\documentclass[a4paper, 12pt]{article}

\usepackage{authblk}
\usepackage[protrusion=true,expansion]{microtype}

\usepackage{graphicx}
\usepackage{subfigure}
\usepackage{times}

\title{Hard and soft bounds in the evolution\\
of Ubuntu packages.\\
A lesson for species body masses?}

\author[1,2]{Marco Gherardi\thanks{Marco.Gherardi@mi.infn.it}}
\author[1]{Salvatore Mandr\`a}
\author[1,2]{Bruno Bassetti}
\author[3,4]{Marco Cosentino Lagomarsino}
\affil[1]{Dipartimento di Fisica, Universit\`a degli Studi di Milano, via Celoria 16, 20133 Milano, Italy}
\affil[2]{Istituto Nazionale di Fisica Nucleare, Sezione di Milano, via Celoria 16, 20133 Milano, Italy}
\affil[3]{Genomic Physics Group, UMR 7238 associ\'ee au Centre National de la Recherche Scientifique,
``Microorganism Genomics,'' 15 rue de l'\'Ecole de M\'edecine, 75006 Paris, France}
\affil[4]{Universit\'e Pierre et Marie Curie, 4 Place Jussieu, 75005 Paris, France}

\begin{document}

\maketitle

Understanding the patterns of software evolution has a large practical
importance: the knowledge of what can be considered ``typical'' can
guide developers and engineers in recognizing and reacting to abnormal
behavior.  While the initial framework of a theory of software
exists~\cite{Lehman2003,softwareevolution}, the current theoretical
achievements do not fully capture existing quantitative data or
predict future trends.  Programs are embedded in the real world, and
consequently the growth of a software package is characterized by
inherent adaptive change in response to complex factors. The
multi-level feedback where programs and their environment evolve in
concert is elusive and difficult to describe precisely.

These very features make the subject attractive from the point of view
of ``complex systems'' theory and analysis.
Most of the ``traditional'' analyses concerned proprietary software,
but a number of studies carried out within the past 10-15 years
gathered a relevant amount of evidence concerning the evolution of
Open Source Software
(OSS)~\cite{Fernandez-Ramil2008,Ermann2011,Maillart2008,Godfrey2000}. The
open source phenomenon has two specificities that make it particularly
interesting.  First, the goal of an open source project is to create a
system that is useful or interesting to its developers, and thus fills
a ``social void'' rather than a commercial one. Second, large OSS
projects are developed and maintained in a globally decentralized
context, contrary to traditional software. The emergent complex
self-organizing structure challenges traditional theories of
management and
engineering~\cite{Madey_Freeh_Tynan_2002,Fortuna21112011}. The OSS
phenomenon is also affecting the daily lives of increasingly many
people, since OSS operating systems and applications run on devices
ranging from PCs to mobile phones and tablets.

Perhaps the simplest observable related to software growth is its
size, which can be measured with different
approaches~\cite{Kemerer1999}.  Despite its simplicity, the size of a
piece of software encapsulates many of the features of its evolution
and evolvability.  Here, we consider the dynamics of package size in a
widely used GNU/Linux system, the Debian-based Ubuntu distribution
(\texttt{www.ubuntu.com/project}).  We analyze systematically the
available data and show that they are compatible with a multiplicative
anomalous diffusion process.  We study this process with the aid of a
theoretical model, and show that the combination of a ``hard'' lower
cutoff and a ``soft'' upper cutoff on package size reproduces with
extreme accuracy the observed distribution.  The same model makes
definite quantitative predictions for the \emph{future} dynamics of
Ubuntu packages.  Finally, as we will see, the knowledge of these
evolutionary patterns might lend a fresh perspective to the debate on
the quantitative aspects of an a \emph{a priori} unrelated process,
the cladogenesis that determines the mass distribution of mammals.

Ubuntu ``packages'' are bundled files comprising the pieces of
software that make up the whole system.  Since Ubuntu was first
released in October 2004, the number of packages increased from a few
hundred to tens of thousands.  Since then, one new release every six
months has been issued. This chronological regularity is valuable for
a systematic quantitative study.
The first, second, and third
releases were christened \emph{Warty Warthog}, \emph{Hoary Hedgehog},
and \emph{Breezy Badger}; from then on, the naming followed
alphabetical order, encompassing $17$ different real and imaginary
animals, up to the latest \emph{Quantal Quetzal} (October 2012).

\begin{figure}[h]
\centering
\subfigure{
\includegraphics[scale=0.9]{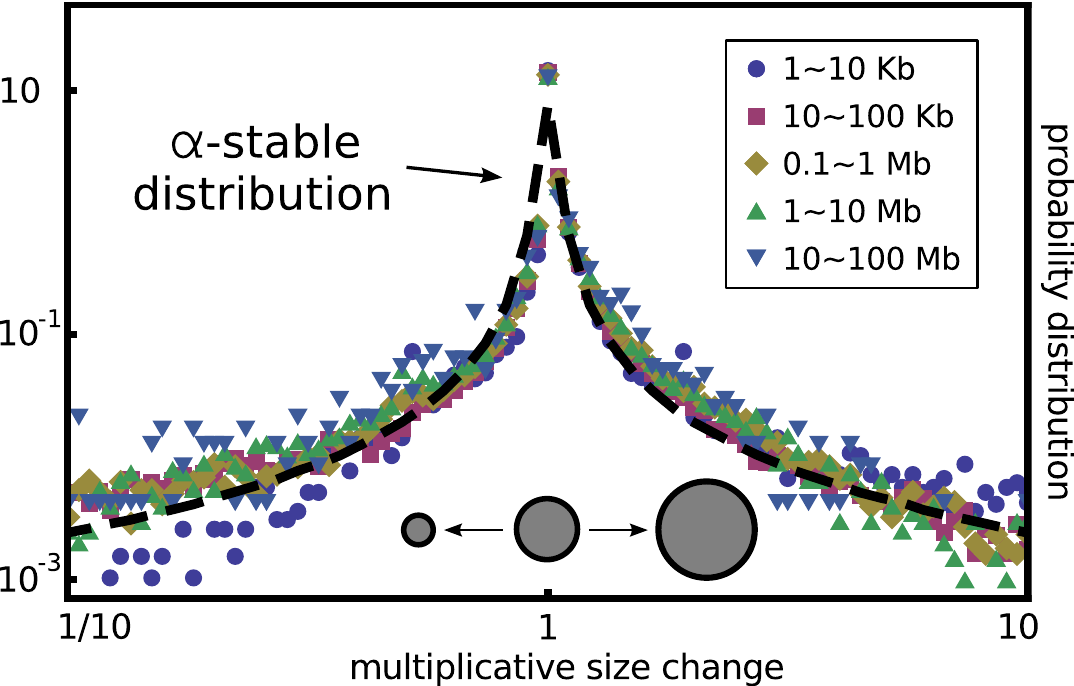}
}
\subfigure{
\includegraphics[scale=0.9]{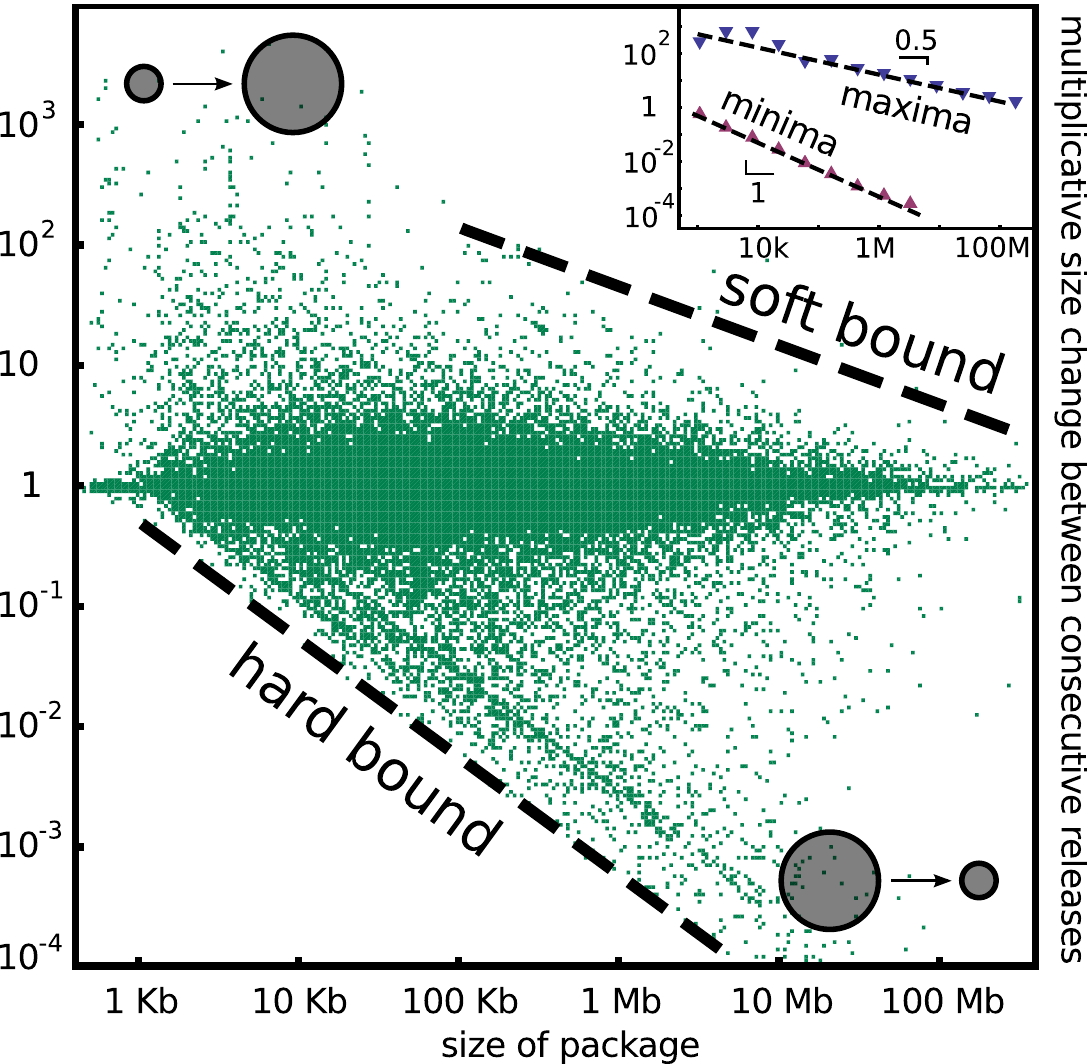}
}
\caption{ \textbf{The changes in package size between Ubuntu releases
    follow an $\alpha$-stable distribution with size-dependent
    bounds.}  (Top panel) The distribution is independent of size for
  small multiplicative size changes ($x$ axis; symbols represent
  different size ranges); (bottom panel) a scatterplot of
  multiplicative size change vs initial package size in the whole
  range reveals a hard lower bound, due to a minimum attainable size,
  and a soft upper bound, with a nontrivial size dependence; the inset
  shows binned averages of maximum and minimum size changes (dashed
  lines are power-law fits yielding exponents $0.5$ and $1$
  respectively).  }
\label{figure:ubuntu_jumps}
\end{figure}

Analysis of empirical data for approximately $370\,000$ changes in
package size, between successive releases
spanning the entire lifetime of Ubuntu, 
reveals striking regularity (Fig.~\ref{figure:ubuntu_jumps}, top
panel).  The logarithm of the multiplicative change $\delta=s'/s$
between the sizes $s$ and $s'$ of a package in consecutive releases
appears to follow
an \mbox{``$\alpha$-stable''} distribution, independently of the
initial size $s$ and of time (the distribution is centered in
$\delta=1$ and has power-law exponent $\alpha\approx 0.7$).  This
class of distributions, widely used in many modeling
contexts~\cite{Mandelbrot:1982,MantegnaStanley:1995,Brockmann:2006,Barthelemy:2008},
contains the most general probability distributions followed by the
sum of a large number of independent identically-distributed random
variables; it is therefore a generalization of the Gaussian (which is
recovered for $\alpha=2$).
Such a multiplicative growth is reminiscent of the trends found in
another area of human interactions, the evolution of business firms'
sizes~\cite{Stanley:2005PNAS}.

Notably, while the bulk of the empirical distribution of
multiplicative size changes is symmetric, events belonging to the
tails are bounded in a size-dependent way
(Fig.~\ref{figure:ubuntu_jumps}, bottom panel).  No package can shrink
to sizes smaller than a global cutoff $s_\mathrm{min}$. This
\emph{hard} bound is easily rationalized by the existence of minimum
requirements from the package management system.  Consequently, the
largest possible decrease is $\delta_\mathrm{min}=s_\mathrm{min}/s$;
from the empirical data we fix $s_\mathrm{min}=741$~bytes, which is
the average of the minimum package sizes found in each release.
Such small packages are usually ``dummy'' packages, not containing any
functional software, but typically pointers to other packages.
Expansion to larger package sizes manifests a more intriguing and
complex behavior: the largest size that a package can attain between
two consecutive releases \emph{depends on its starting size}.
Specifically, the largest possible increase is
$\delta_\mathrm{max}=(s_\mathrm{max}/s)^\gamma$, with an exponent
$\gamma$ approximately equal to $1/2$. We call this a \emph{soft}
bound, meaning that larger packages are less prone to perform large
jumps, but packages of different initial sizes do not behave as if a
unique maximal size were present.  Trade-offs between increase in
complexity and cost of deployment are probably responsible for this
non-linear law.  Birth of new packages is also a relevant driving
process for the dynamics of software evolution.  Newborn packages
appear with size proportions approximately equal to those of the
preceding release,
which suggests that the dominant route of expansion is software
forking and reuse (but note that the data set does not distinguish new
packages from old packages with new names).

Based on the foregoing empirical observations, we define a stochastic
model of package size evolution, which relies on three assumptions:
(i) At every new release, each package (of size $s$) assumes the new
size $s'=s\delta$ (multiplicative size changes).
(ii) Each package has a small probability $p$ of also adding a copy of
itself to the new release (branching).
(iii) The logarithms of the growth factors $\delta$ are independent
$\alpha$-stable random variables conditioned on two size-dependent
cutoffs, a lower hard bound and an upper soft bound, as evidenced by
the data.  Technically, this model is realized as a branching
multiplicative diffusion process. We do not explicitly consider
package deletion, which does happen in Ubuntu; however, we found that
its role is irrelevant for the evolution of package size
distributions.  The model above has no free parameters, as all the
quantities needed to specify the distribution are estimated by data
analysis.

\begin{figure}[h]
\centering
\includegraphics[scale=1]{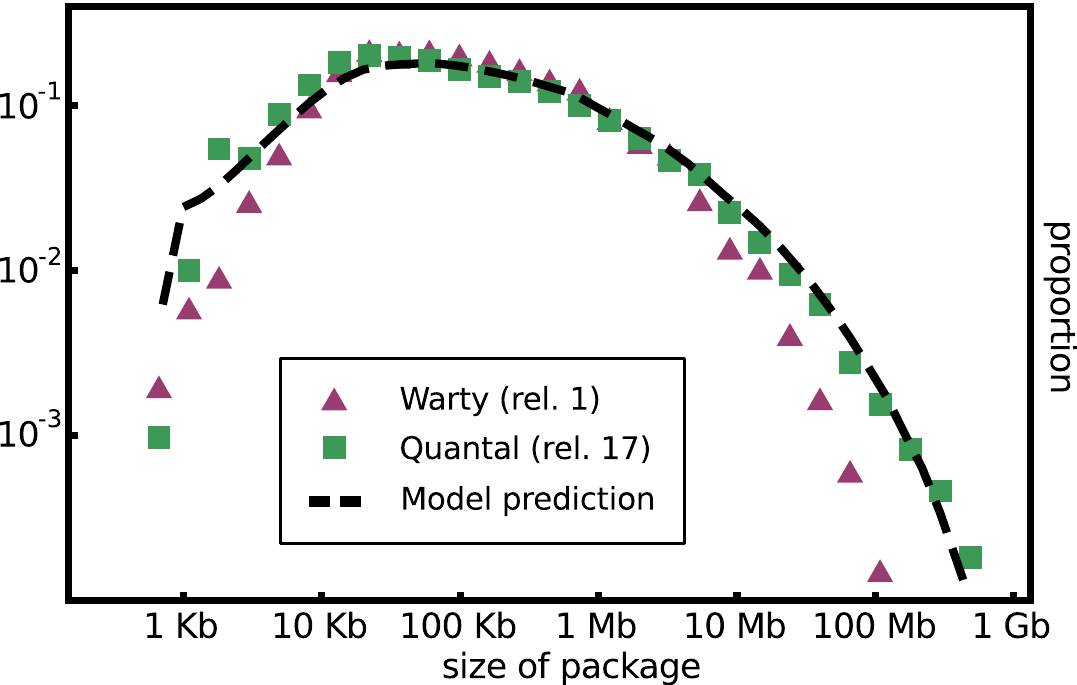}
\caption{ \textbf{The dynamics of package-size distribution is
    captured by a branching multiplicative diffusion process.}
  Starting from the initial pool of packages constituting \emph{Warty
    Warthog} (triangles), with all parameters fixed by data analysis,
  the model yields the distribution traced by the dashed line, which
  nicely reproduces size proportions in \emph{Quantal Quetzal}
  (squares).  Notice that the tails of the two empirical distributions
  differ by almost one order of magnitude; furthermore, the ramp at
  small sizes for \emph{Quantal} (which was not present in
  \emph{Warty}) is correctly predicted.  }
\label{figure:ubuntu_distribution}
\end{figure}

Starting from the population of packages in the first Ubuntu release,
\emph{Warty}, and evolving their sizes for $16$ steps (eigth years),
the model predicts very accurately the package size distribution in
the latest release, \emph{Quantal}
(Fig.~\ref{figure:ubuntu_distribution}).  Sensitivity analysis shows
that the results are robust with respect to variation of the
parameters.  Moreover, the accordance of model and data is not
dependent on the particular shape of the distribution; in fact,
arbitrarily chosen subsets of packages can be followed through their
evolution, and the size proportions they assume in \emph{Quantal} are
reproduced strikingly well.
The model is predictive, and can be used to forecast future
evolution. For instance, we found that the current distribution is
very far from stationary; at this rate, a stationary state would be
reached in approximately $2$--$400$ years.
In $10$ years the largest package should weigh approximately $1$ Gb,
and the average package size is predicted to nearly double from the
current $1.2$ Mb to about $2.3$ Mb; the most common size, instead,
will have slightly increased only by around $10$ kb (it is currently
$22$ kb).

We found that the knowledge of the anomalous diffusion framework with
``soft'' cutoffs described above may suggest a different
perspective on the debate around a distant scientific problem.  
In fact, similar models to the one described here have been employed to 
explain the evolution of species body masses in mammals and other
taxa~\cite{Alroy1998,Clauset2008}.
In this case, the branching process represents cladogenesis, i.e.\ the
lineage splitting event generating new species (\emph{clades} in the
phylogenetic tree) whose average body mass is related to the
ancestor's.  The model proposed by Clauset and
Erwin~\cite{Clauset2008} and further developed in
\cite{ClausetRedner:2009,ClausetSchwab:2009} assumes multiplicative
diffusion on evolutionary time scales, with a lower hard bound due
to metabolic constraints, and an explicit bias toward larger sizes
(the controversial ``Cope's rule''
\cite{Cope:1887,Gould:1997,Valkenburgh:2004,Moen:2006}),
whose strength must increase for lower masses (although there appears to be
also evidence for the opposite tendency \cite{Alroy1998}).  Moreover, the
introduction of a size-dependent extinction rate is necessary in order to
approximate the large-mass tail of the empirical distribution of
extant mammals.

\begin{figure}[h]
\centering
\subfigure{
\includegraphics[scale=1]{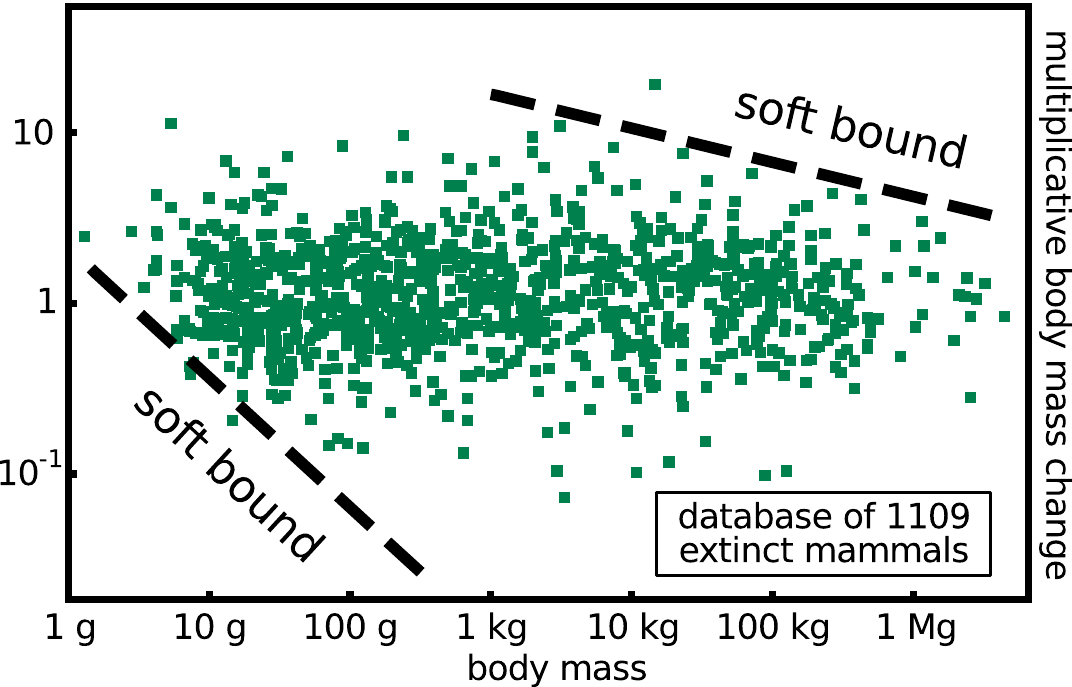}
}
\subfigure{
\includegraphics[scale=1]{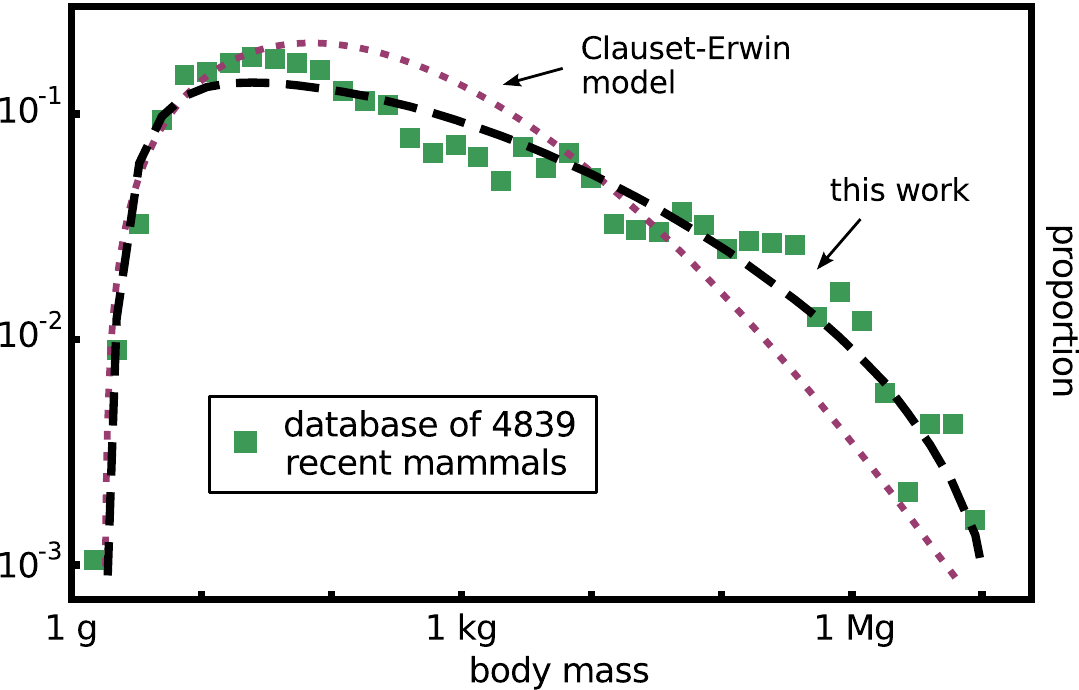}
}
\caption{ \textbf{ Application of the bounded diffusion framework to
    mammalian body-mass data.}  (Top panel) Multiplicative changes in
  body mass for $1109$ mammalian species, plotted as a function of
  ancestor's body mass (squares); data are compatible with the
  existence of soft bounds (dashed lines), but do not allow to define
  them.  (Bottom panel) The distribution of mammalian body masses is
  well reproduced by the model (dashed line), and some features appear
  to be improved with respect to the Clauset-Erwin model (dotted
  line).  Note that the dotted line corresponds to a stationary state,
  while the dashed line does not.  }
\label{figure:mammals}
\end{figure}

In the framework suggested by software evolution, it seems natural to
characterize the low propensity of large species to generate
exceedingly large descendant species (and the tendency of small
species to generate larger ones) through a ``soft'' cutoff instead.
Fossil data of ancestor-descendant size ratios are not abundant, and
susceptible to noise and bias~\cite{Liow:2008}.  We used
a compilation by Alroy of $1109$ North American terrestrial mammals up
to the late Pleistocene, obtained by a highly conservative method
\cite{Alroy1998}.  Despite the great amount of work behind these data,
they do not allow an estimate of parameters nearly as precise as what
was attained for Ubuntu packages; nonetheless,
our analysis shows that
the changes in body size are compatible with upper and lower soft
cutoffs with $\gamma$-values around $0.2$ and $0.6$ respectively 
(see Fig.~\ref{figure:mammals}, top panel).
Uncertainties on these estimates are not a big inconvenience, as the
results are fairly robust to variation of these parameters.

We simulated the \emph{in silico} evolution of body masses throughout
mammalian history, starting from the body mass of the founder species
\emph{Hadrocodium wui}, a small mammaliaform from the Early Jurassic
weighing $2$ grams~\cite{Luo:2001}. Remarkably, the characteristically
skew and wide distribution of extant mammals \cite{MOM} is recovered
with good precision by this model (Fig.~\ref{figure:mammals}, bottom
panel).  The (softly) bounded nature of the diffusion, together with
the asymmetry of the initial condition, are the key ingredients that
account for the shape of the empirical distribution.
It must be said that the agreement is not completely parameter-free as
in the case of Ubuntu packages: model time is chosen as the one that
best recovers the expected distribution, since it cannot be estimated
directly. However, one or more free parameters were
present also in all previous studies~\cite{Clauset2008,ClausetRedner:2009}.

One important remark is that the present model relaxes the common
assumption that the mammalian body-mass distribution is stationary at
present time.  Consequently, different initial conditions can produce
markedly different distributions.  If the initial mass is sufficiently
large, left-skewed distributions can be obtained; such a shape is less
common but it is nonetheless found in some taxa~\cite{Kozlowski:2002}.
Mining the literature, we could not find any strong evidence either
supporting or undermining the assumption of stationarity, and
therefore we hope that our findings could be useful to stimulate the
debate in this direction.  Incidentally, we note that a further
prediction of the model is a slowly saturating evolution for the
maximum body mass as a function of time, which is in line
with recent findings~\cite{Smith:2010}.

A second and final remak is that the bounds on the diffusion process
in the context of mammalian body masses are realized by a
size-dependent extinction rate~\cite{Liow:2008}.  In our approach, the
soft nature of the constraint for large masses is interpreted as the
result of the competition between the short-term selective advantages
of an increased body size and the corresponding long-term extinction
risk, as concluded in previous studies.  This macroevolutionary
tradeoff mechanism is quantitatively robust across all mammalian
species \cite{Clauset:2013}, and also in other taxa
\cite{ClausetSchwab:2009}.  The observation that the lower boundary is
soft as well suggests that a similar tradeoff may be present also for
small body masses.

\paragraph*{Acknowledgements}
We are grateful to Aaron Clauset for help with the Alroy dataset.  We
wish to thank Alberto Vailati, Vincenzo Gino Benza, and Matteo Osella
for helpful suggestions.

\bibliographystyle{unsrt}
\bibliography{ubuntu_mammals}

\end{document}